\documentclass[%
 aip,
 amsmath,amssymb,
reprint,%
]{revtex4-2}
\newcommand{\ra}{\rangle}

\usepackage{graphicx}
\usepackage{epsfig}
\begin{document}
\title{Effects of dissipation in reservoir computing using spin qubit array}

\author{Shion Mifune}
\affiliation{Department of Information and Electronic Engineering, Teikyo University,
Toyosatodai, Utsunomiya 320-8511, Japan} 

\author{Taro Kanao}

\affiliation{Computer Science and Engineering Course, 
College of Engineering, Shibaura Institute of Technology, 
Toyosu, Koto-ku, Tokyo 135-8548, Japan} 

\author{Tetsufumi Tanamoto}
\affiliation{Department of Information and Electronic Engineering, Teikyo University,
Toyosatodai, Utsunomiya 320-8511, Japan} 

\begin{abstract}
Reservoir computing (RC) is one of the hottest topics as a promising 
application for many physical devices because their device characteristics 
can be directly used in the computing sequences. 
Quantum RC is also the promising candidate for the application of 
even small-number qubit system. 
Here, we propose a quantum RC based on spin qubit system reflecting 
the status of the spin qubits in experiments of one-dimensional qubit array.
Spin qubits are coupled via the Heisenberg interaction, and data sequences are 
inputted to one of the spin qubits by pulsed rotations.
By introducing dissipation, we obtain satisfactory performance as a quantum RC.
\end{abstract}

\maketitle
%
%
%
%
%

\section{Introduction}
Spin qubits are one of the promising platform of quantum computing devices.
Since the proposal of Loss and DiVincenzo~\cite{Loss}, 
many theories and experiments have been carried out so far. 
In the early stage of the experiments, GaAs systems were used, and at present, 
Si based spin qubits have been the main stream of experiments aiming 
at the seamless fusion to the conventional silicon computers.
The target architecture of qubits is the surface code consisting of two-dimensional 
qubit array~\cite{Fowler},
and many proposals have been made~\cite{Hill,Veldhorst1,tanaJAP}.
However, the state-of-the-art experiments of Si qubits are one-dimensional qubit array~\cite{Intel,Dzurak,Vandersypen,Tarucha1}.
Although these are great progress from the early stage, 
this status of the spin qubits is in contrast with the developments of the superconducting qubits,
where advanced institute and universities are competing 
with increasing the number of qubits in surface code~\cite{Google2023,IBM2023}.
The advantage of the superconducting qubits is their longer coherence time 
even though the size of the qubits is large.
On the other hand, the traditional spin qubits require many wires 
for the control and measurement ('jungle of wiring problem').
This problem may be solved when the advanced 2 nm fabrication 
process technologies are spread widely in commercial phase. 
We believe that, until that time, some kind of realistic application 
might be needed to upgrade the spin qubit technologies.
In this study, we investigate the application of 
the reservoir computing using the spin qubit array.
The reservoir computing (RC) is one of the promising recurrent neural networks~\cite{ESN,Maass,Verstraeten}.
Compared with general recurrent neural network, 
the feedbacks of weight functions are limited to the output part of the network.
Thus, the speed of learning is larger than the conventional recurrent neural network.
In addition, various natural physical systems can be directly used as it is 
to the network, therefore various proposals have been made~
\cite{AIST,Nakane,Takagi,Nakajima1,Kanao}.

According with the development of classical RC mentioned above, 
quantum RCs (QRCs) also have been proposed by many researchers~\cite{Fujii,Fujii2,Yamamoto1,Yamamoto2}.
Fujii and Nakajima~\cite{Fujii,Fujii2} proved the effectiveness of the QRC
in the nuclear magnetic resonance (NMR) system where the Ising interaction is used.
However, we think that their approaches do not directly apply to the solid-state 
silicon qubits from two points. 
The first problem is regarding how to input the data sequence into the qubit state.
Fujii and Nakajima~\cite{Fujii,Fujii2} and many other researchers 
determine the input qubit state such that 
the coefficients of the wave function directly reflect the input data (0 or 1) 
by using the completely positive and trace preserving (CPTP) map.
But, it is not easy to determine the qubit state in spin qubits 
each time of the input data sequence.

The second problem is regarding the controlled-NOT (CNOT) operation.
Yamamoto's group has proposed a RC for superconducting qubit 
where operations between qubits are CNOT operations~\cite{Yamamoto1,Yamamoto2}.
However, in spin qubits, at present, 
the CNOT gate operations are still "expensive"\cite{Veldhorst,Tarucha2,Petta}.
Because in RC, a lot of operations are required,
it is not realistic to use many CNOT gates.
In this respect, we have to find out the appropriate input method that is fit with 
the spin qubits.
 
In this study, we investigate the QRC approach which is in line with 
the status of the spin qubit experiments. 
We consider the spin qubit RC using only pulse sequences
starting from a fixed initial state.
The input data including 0 and 1 are regarded as the 
rotation angle $\theta_k$  of the $x$-axis rotation.
Instead of CNOT gates, we utilize always-on Heisenberg interactions,
which can be naturally induced in spin qubit systems.

 In an echo-state network (ESN)\cite{ESN}, the reservoir internal state $x_k$ at the time 
step $k$ is a function of the $x_k$ and its previous data $x_{k-1}$.
This means that the status at the time step $k$ does not explicitly include the previous data 
$x_{k-m}$ for $m>1$.
This setting can lead to the important behavior of the ESN called echo state property.
However, in the general physical system, the state at $k$ also depends on 
those of $x_{k-m}$ for $m>1$. 
Thus, we also estimate the ESN where the time evolution 
includes previous states more than two steps before.

 In Ref.~\cite{SSDM}, we have shown basic results of our theory. 
 In this study, we have extended the theory and newly
 added Figs.~\ref{fig3b}, \ref{fig4a}, \ref{fig4b} and \ref{PSB}.
The remainder of this paper is organized as follows. 
The basic formulation of RC in spin qubit is presented in Section 2. 
Section 3 provides the consideration regarding dissipation. 
Section 4 presents the numerical results. 
In Section 5, we discuss the results, and finally, Section 6 summarizes and concludes the study.

\section{Reservoir computing for spin qubit system}
\subsection{Previous approach to QRC}
In Ref.~\cite{Fujii}, the data sequences are inputted into the first qubit 
which is separated from other part (input encoding method).
In this method,
for the input bit sequence $s_k$ ($s_k \in \{0,1\}$),
the input qubit is defined and its quantum state $|\Psi_{s_k}\rangle^{\rm in} $is 
changed into 
\begin{equation}
|\Psi_{s_k}\ra^{\rm in}=\sqrt{1-s_k}|0\ra +\sqrt{s_k} |1\ra.
\label{cptp}
\end{equation}
Then, the system density matrix is transferred to
\begin{equation} 
\rho \rightarrow \rho_{s_k} \otimes {\rm Tr}_{\rm in}\rho,
\end{equation}
where Tr$_{\rm in}$ shows the trace with respect to the input qubit.
The measurement and feedback are required to form the wavefunction to Eq.(\ref{cptp}) 
(CPTP mapping).
However, 
it is not clear to how to satisfy the CPTP map in the noisy solid-state spin qubit system.

The density matrix $\rho_{k+1}$ at time step $k+1$ is given by 
$\rho_{k+1}=\rho_{\rm in} \otimes {\rm Tr}_1 (\rho_k)$ 
where $\rho_{in}$ is the density matrix of the input qubit, 
and ${\rm Tr}_1$ indicates the partial trace regarding the input qubit.
This scheme showed excellent RC performance. 
However,  
in case of the spin qubits~\cite{Tarucha2}, spins interact with always-on interaction, 
and they are controlled only by the external pulsing magnetic field.
Thus, it is not easy to separately treat the input qubit 
from other qubits.
Therefore, the input encoding method is not realistic 
for spin qubits.

\begin{figure}
\centering
\includegraphics[width=6cm]{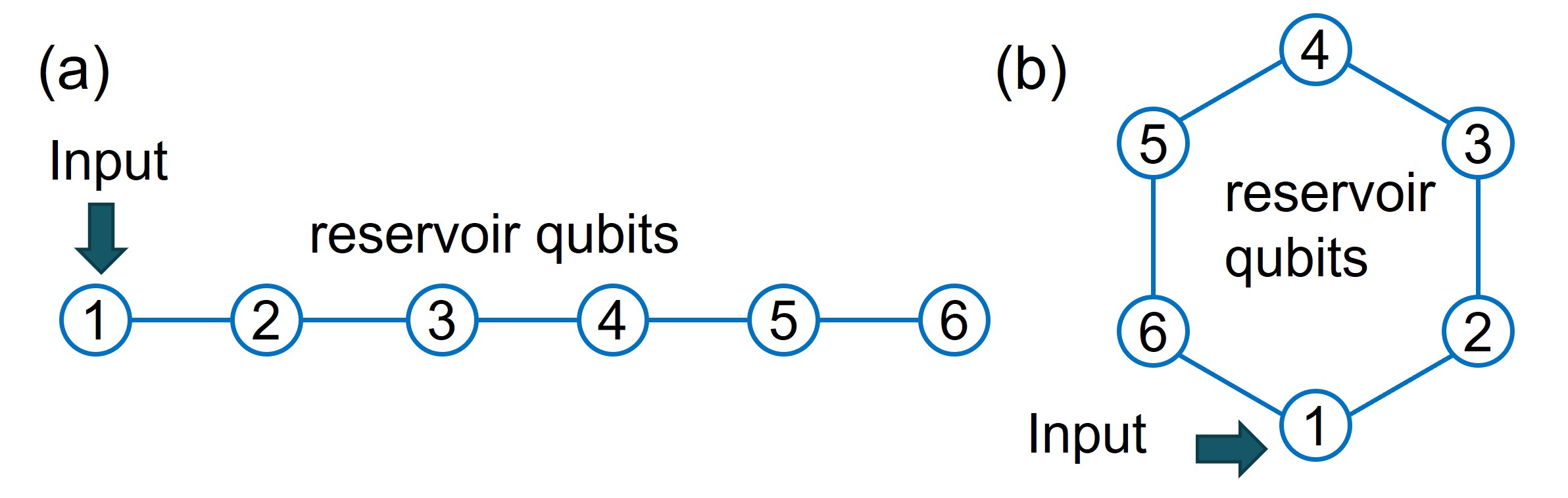}
\caption{
Reservoir system using six spin qubits ($N_q=6$).
(a) Linear arrangement. (b) Ring arrangement.
The numbered circles correspond to
the spin qubits. 
}
\label{fig1}
\end{figure}
\subsection{Our approach to QRC using spin qubits}
Figure 1 shows the two arrangement of spin qubit system.
We consider the spin qubits system which is interacting by 
Heisenberg interactions whose Hamiltonian is given by
\begin{equation}
H_{\rm int}=\sum_{i,j}^{N_q} J_{ij}( X_iX_j + Y_iY_j+Z_iZ_j),
\end{equation}
where $X_i$, $Y_i$, and $Z_i$ are the Pauli matrices given by
\begin{equation}
X=
\left(
\begin{array}{cc}
0 & 1 \\
1 & 0
\end{array}
\right), \ 
Y=
\left(
\begin{array}{cc}
0 & -i \\
i & 0
\end{array}
\right), \ 
Z=
\left(
\begin{array}{cc}
1 & 0 \\
0 & -1
\end{array}
\right), 
\end{equation}
(we set $\hbar=1$ and the number of the qubits is $N_q=6$). 
By solving the eigenvalue problem,
the wave function is given by $|\psi_k\ra$.
The unitary matrix is given by
\begin{equation}
U_{\rm int}(t)=\exp (-i t H_{\rm int}).
\end{equation}
In the following calculations, $J_{ij}$ is randomly taken 
such as $0\le J_{ij} \le 1$ (we set the energy-scale by 
max($J_{ij})=1$).
It is assumed that a sufficiently strong static magnetic field, relative to the operating temperature, is applied. 
Each electron can exist in two spin states: spin-up and spin-down, due to Zeeman splitting, which represents a two-level system
such as Ref.\cite{Intel,Dzurak,Vandersypen,Tarucha1}.
The input data sequence $s_k$ 
is taken in as the rotation angle around $X$ axis
given by, 
\begin{equation}
R_X(s_k)=\exp \left( i\pi s_k X_1/2 \right),
\end{equation}
(the rotating wave approximation is assumed)\cite{Nakajima2}.
The system evolves by the unitary operator $U(t)$ given by
$
U(t)=\Pi_k R_X(s_k) U_k 
$
where
$
U_k = \exp (-i\Delta t H_{\rm int})
$
with $\Delta t=\pi \theta_0$.
Although we tried values of $\theta_0$, we typically take 
$\theta_0=0.5$ in the following calculations.

We assume the $Z$-measurement is carried out by the measurement. 
\begin{equation}
\langle Z_i\rangle = {\rm Tr} Z_i \rho (t).
\end{equation}
The direct measurement of reservoir qubits determines their state as 0 or 1,
therefore, some ensemble-averaging process is required (see discussion below).
  
The entire process consists of $N_{\rm pre}$ 
preparation steps, $N_{\rm fb}$ learning steps, and $N_{\rm test}$ testing steps
(here, $N_{\rm pre}=N_{\rm fb}=200$, and $N_{\rm test}=40$).
There are a couple of choices to pick out the reservoir outputs.
Lee and Mochizuki~\cite{Lee} showed that the 
$Z$-component magnetization of skyrmion averaged over sites
shows satisfactory performance. 
Direct usage of the six qubit measurement is another candidate.
Thus, we consider two types of relations between the measurement results 
and the reservoir outputs ($y_k^{\rm I}$ and $y_k^{\rm II}$).
In the first type  $y_k^{\rm I}$,  
$N_q$ measurement results 
$
x_{ki}= \langle Z_i \rangle
$
are used and given by
$ 
y_k^{\rm I}=\sum_{i=0}^{N_q} x_{ki}w_i,
$ 
where $w_i$ is the readout weights ($i=0,..,N_q$), and $x_0=1$ is a constant.
In the second type, 
the average measurement result is 
used  such as 
$
x_k=\frac{1}{N_q}\sum_{i=1}^{N_q} \langle Z_i \rangle
$, 
and $y_k^{\rm II}$ is given by
$
y_k^{\rm II}=x_{0}w_0+x_{k}w_1,
$
where $x_{0}=1$ and the $w_i$ is the readout weight.
Both reservoir outputs are determined by minimizing 
$\sum_{k} |y_k-\bar{y}_k|^2$
for the target output $\bar{y}$.
In the test phase, $y_k$ is predicted by
$y_k^{\rm pred}=\sum_i x_{ki}w_i$
for $x_{ki}$ in the test region.

Here, the short-term memory (STM) task
and the nonlinear autoregressive moving average (NARMA) task are studied.
The STM tests the ability to retain a small amount of data for a brief period.
The STM task is evaluated as the time delay $\tau_B$ given by 
\begin{equation}
y_k=s_{k-\tau_B},
\end{equation}
which characterizes the ability of short-term memory~\cite{Fujii}.
The NARMA task evaluates how the output of dynamical systems relies on both previous output and input history.
The input of the NARMA task is given by
\small
\begin{equation}
s_k =0.1
\left[
\sin \left (\frac{2\pi\alpha_0 k}{T}\right)
\sin \left (\frac{2\pi\alpha_1 k}{T}\right)
\sin \left (\frac{2\pi\alpha_2 k}{T}\right)+1
\right],
\end{equation}
\normalsize
where $(\alpha_0,\alpha_1,\alpha_2)
=(2.11,3.73,4.11)$ and $T=100$.
The error is estimated by the standard normalized mean squared error (NMSE)~\cite{Fujii,Kanao} given by
\begin{equation}
{\rm NMSE}
=\frac{\sum_{k=L+1}^{M} (\bar{y}_{k+1}-y_{k+1})^2}
{\sum_{k=L+1}^{M}\bar{y}_{k+1}^2},
\end{equation}
where $L=N_{\rm prep}+N_{\rm fb}$ is the number of the time step in the training phase,
$M=L+N_{\rm test}$, and $N_{\rm test}$ is the number of the time step in the evaluation phase.

\section{Dissipation}
The echo-state property 
requires short-term memory but not long-term memory.
In physical system, this means that dissipation is strong.
Then, we here include dissipation as follows. 
The two qubit states $|0\rangle$ and $|1\rangle$ are defined
as the spin-down and spin-up states, 
assuming that there is a strong $z$-direction magnetic field.
We assume that the dissipations occur by various noises 
and the excited state $|1\rangle$ decays into the ground state 
$|0\rangle$ with a decay rate $\gamma$. 
Then, the evolution of the density matrix can be expressed by
\begin{equation}
\rho_{k+1}= U_k \rho_{k} U_k^{\dagger} (1-\gamma) +\gamma \rho_0,
\end{equation}
where $\rho_0 \equiv |0,...,0\rangle \langle 0,...,0|$.
Although we tried several wave functions, there is no significant difference.
Thus, the following calculations are carried out starting 
from the initial state $|0,...,0\rangle$ at time step 0.

\begin{figure}
\centering
\includegraphics[width=8.8cm]{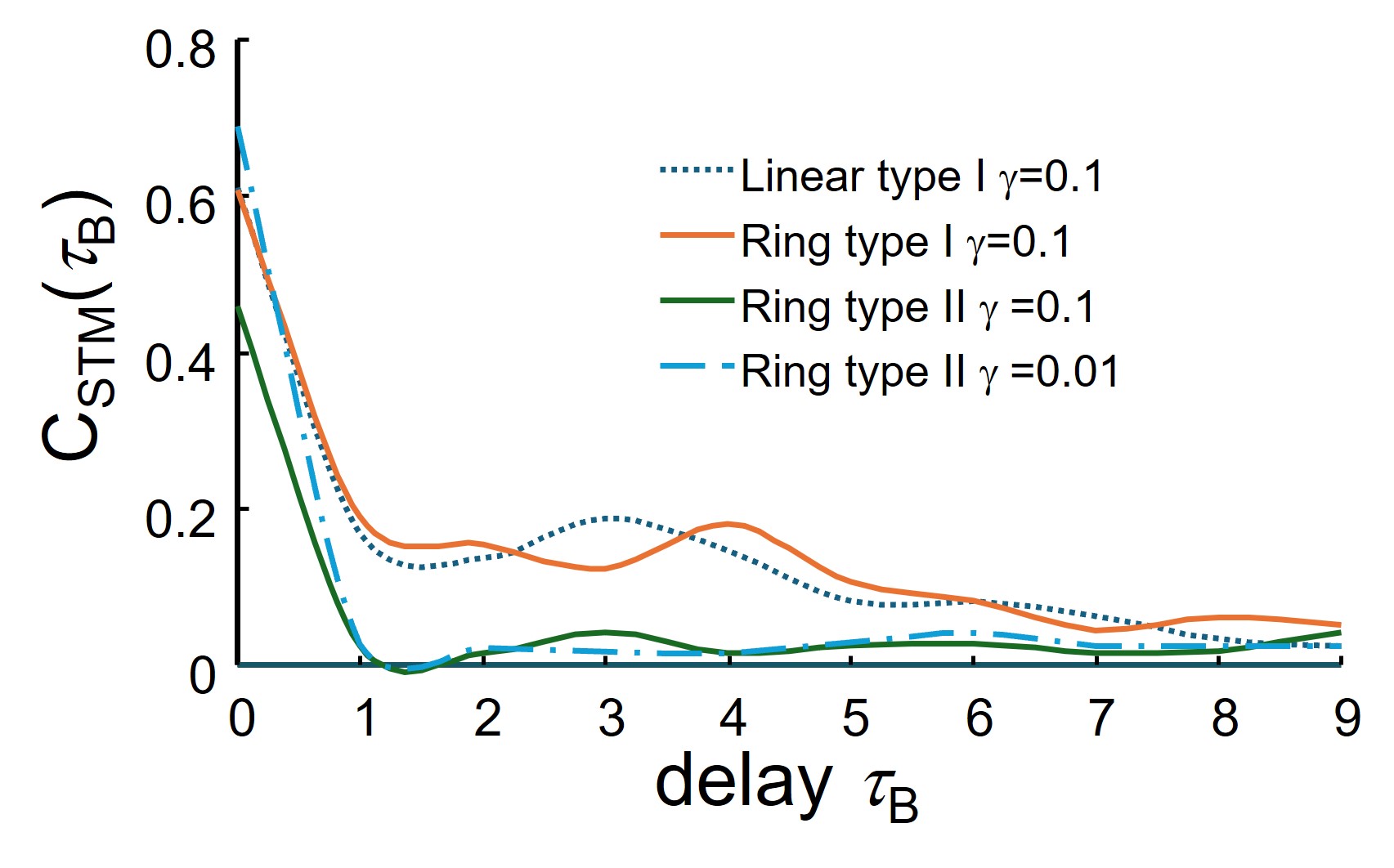}
\caption{
The STM tests for the linear and ring arrangement 
of Fig.1 for $\gamma=0.1$ and $\gamma=0.01$.
}
\label{fig2}
\end{figure}

\small
\begin{table*}
\caption{\label{f1}\label{tabl3}NMSE in NARMA tasks of two $\gamma$s for a type I and type II qubit array.}
\begin{tabular}{lllllllll}
\hline
& \multicolumn{4}{c}{Type I } & \multicolumn{4}{c}{Type II } \\
 \cline{2-9}
    &$\gamma$=0.1&$\gamma$=0.01&$\gamma$=0.1&$\gamma$=0.01&$\gamma$=0.1&$\gamma$=0.01&$\gamma$=0.1&$\gamma$=0.01\\
	\hline
NARMA2 &2.25E-05 &5.76E-05 &2.58E-05 &7.00E-05 &7.45E-06 &5.01E-05 &6.52E-06 &4.01E-05\\
NARMA5 &1.76E-03 &6.17E-03 &1.81E-03 &6.74E-03 &1.81E-03 &6.26E-03 &2.17E-03 &5.66E-03\\
NARMA10&3.65E-03 &6.12E-03 &3.37E-03 &5.88E-03 &4.18E-03 &6.13E-03 &3.88E-03 &5.55E-03\\
NARMA15&4.37E-03 &4.35E-03 &4.05E-03 &4.13E-03 &4.72E-03 &4.76E-03 &4.82E-03 &4.73E-03\\
NARMA20&3.07E-03 &2.83E-03 &3.35E-03 &2.99E-03 &3.13E-03 &2.92E-03 &3.16E-03 &2.87E-03\\
\hline
\end{tabular}
\end{table*}
\normalsize

\begin{figure}
\centering
\includegraphics[width=8.8cm]{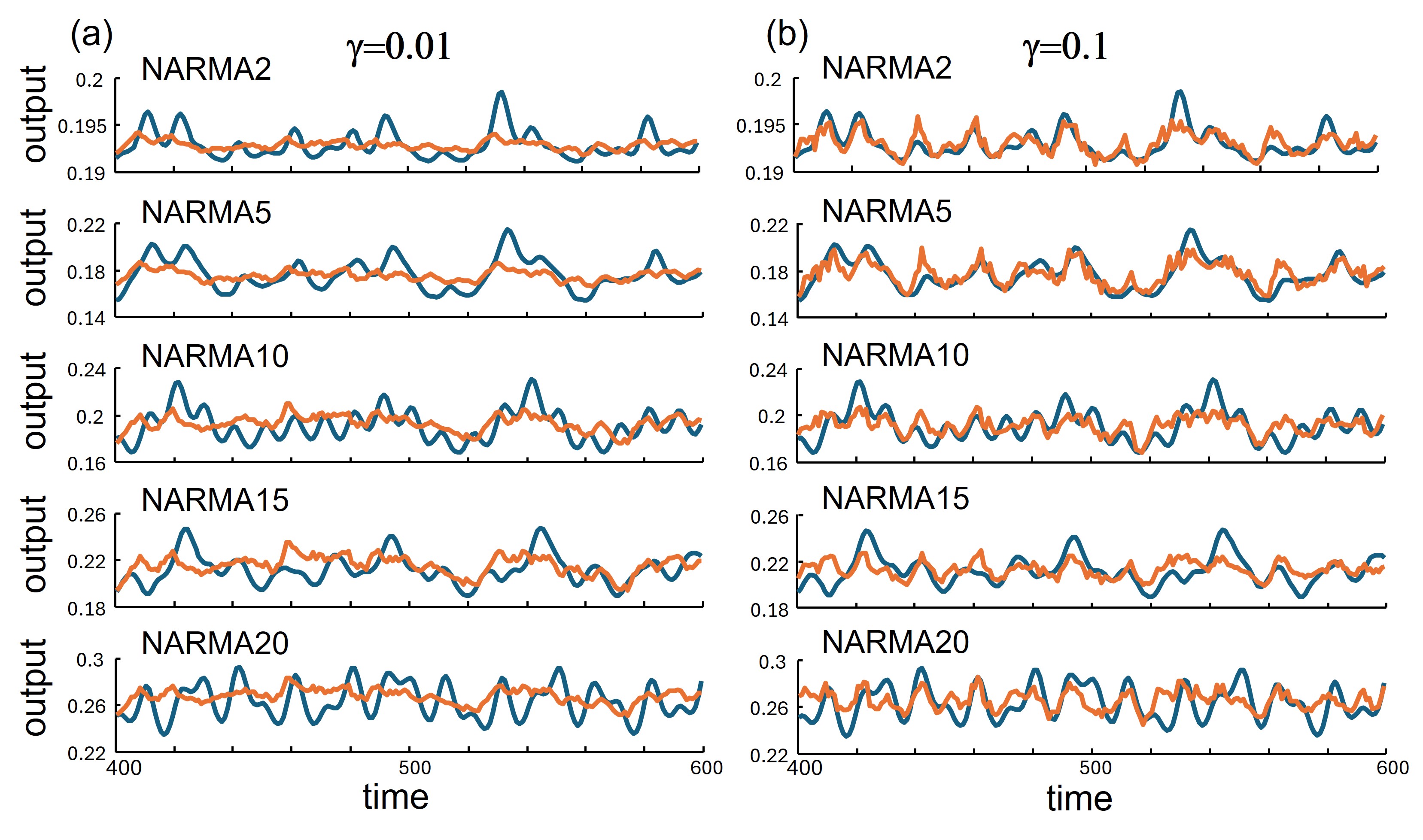}
\caption{A typical example of linear structure (Fig.1(a)) for the NARMA tasks
for $\gamma=0.01$ (left column) and $\gamma=0.1$ (right column) in type I.
The darkcyan and orange lines show the target output and the reservoir output,
respectively.
}
\label{fig3a}
\end{figure}
\begin{figure}
\centering
\includegraphics[width=8.8cm]{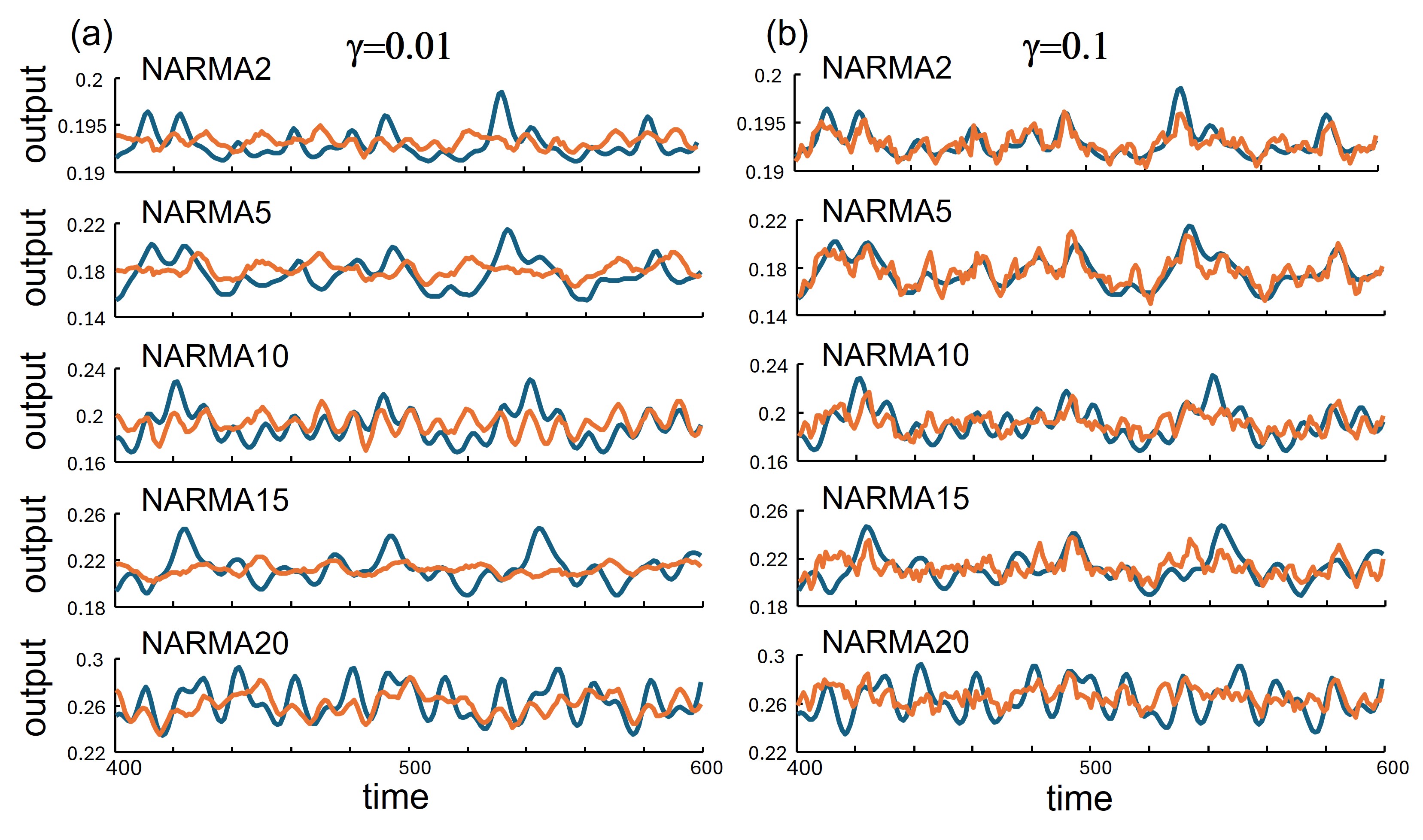}
\caption{A typical example of ring structure (Fig.1(b)) for the NARMA tasks
for $\gamma=0.01$ (left column) and $\gamma=0.1$ (right column)in type I.
The darkcyan and orange lines show the target output and the reservoir output,
respectively.
}
\label{fig3b}
\end{figure}

\begin{figure}
\centering
\includegraphics[width=8.8cm]{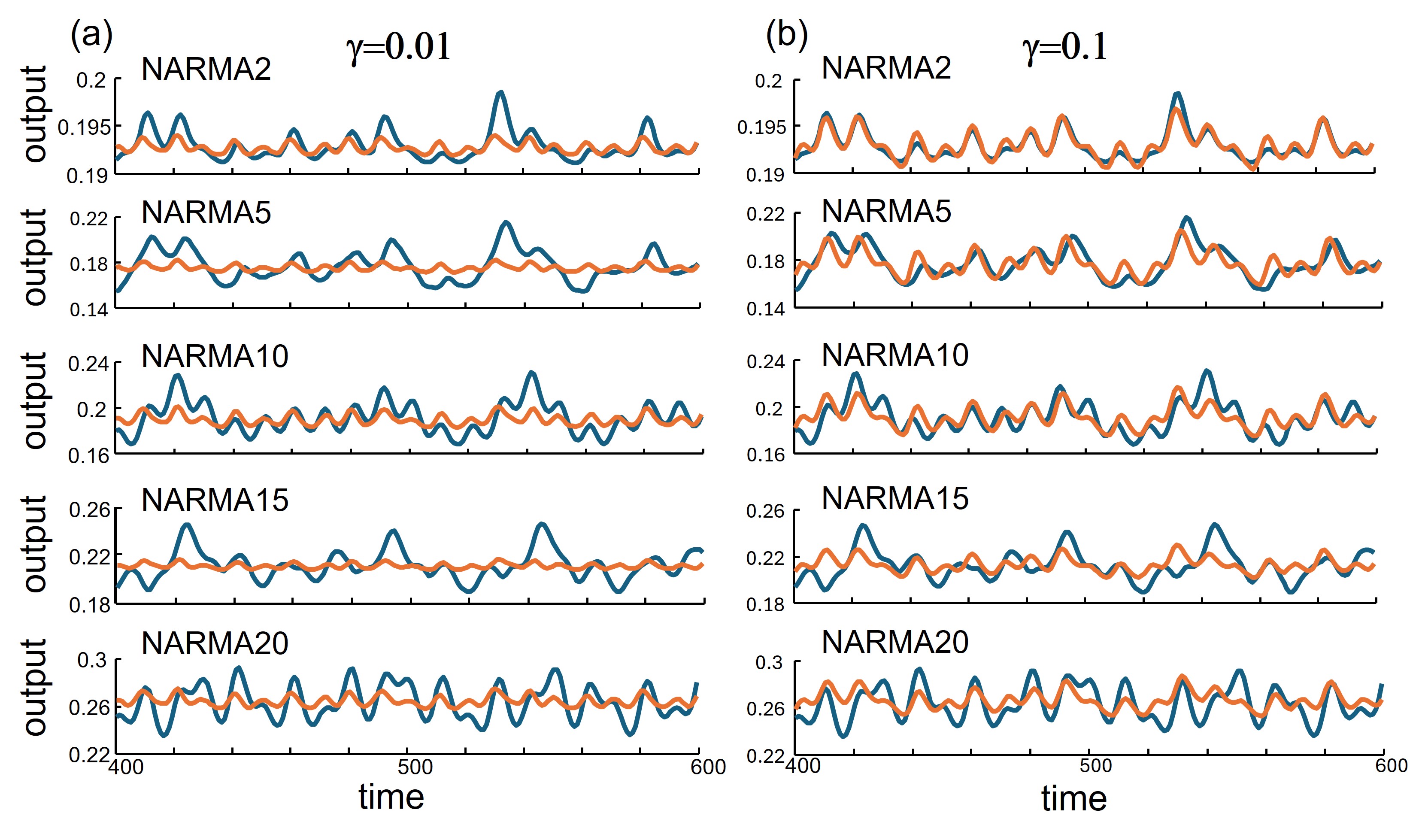}
\caption{A typical example of linear structure (Fig.1(a)) for the NARMA tasks
for $\gamma=0.01$ (left column) and $\gamma=0.1$ (right column) in type II.
The darkcyan and orange lines show the target output and the reservoir output,
respectively.
}
\label{fig4a}
\end{figure}
\begin{figure}
\centering
\includegraphics[width=8.8cm]{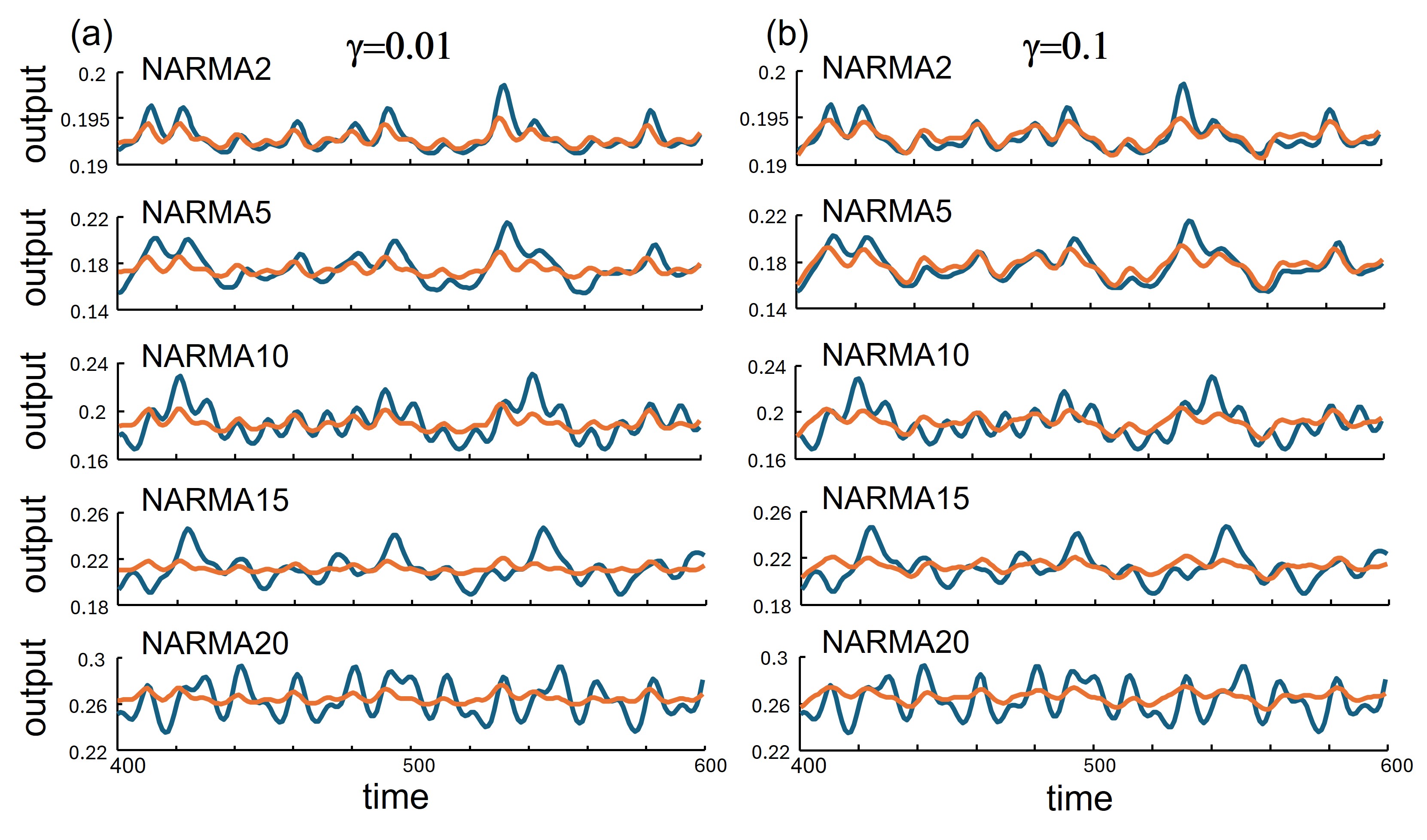}
\caption{A typical example of ring structure (Fig.1(b)) for the NARMA tasks
for $\gamma=0.01$ (left column) and $\gamma=0.1$ (right column) in type II.
The darkcyan and orange lines show the target output and the reservoir output,
respectively.
}
\label{fig4b}
\end{figure}
\begin{figure}
\centering
\includegraphics[width=8.2cm]{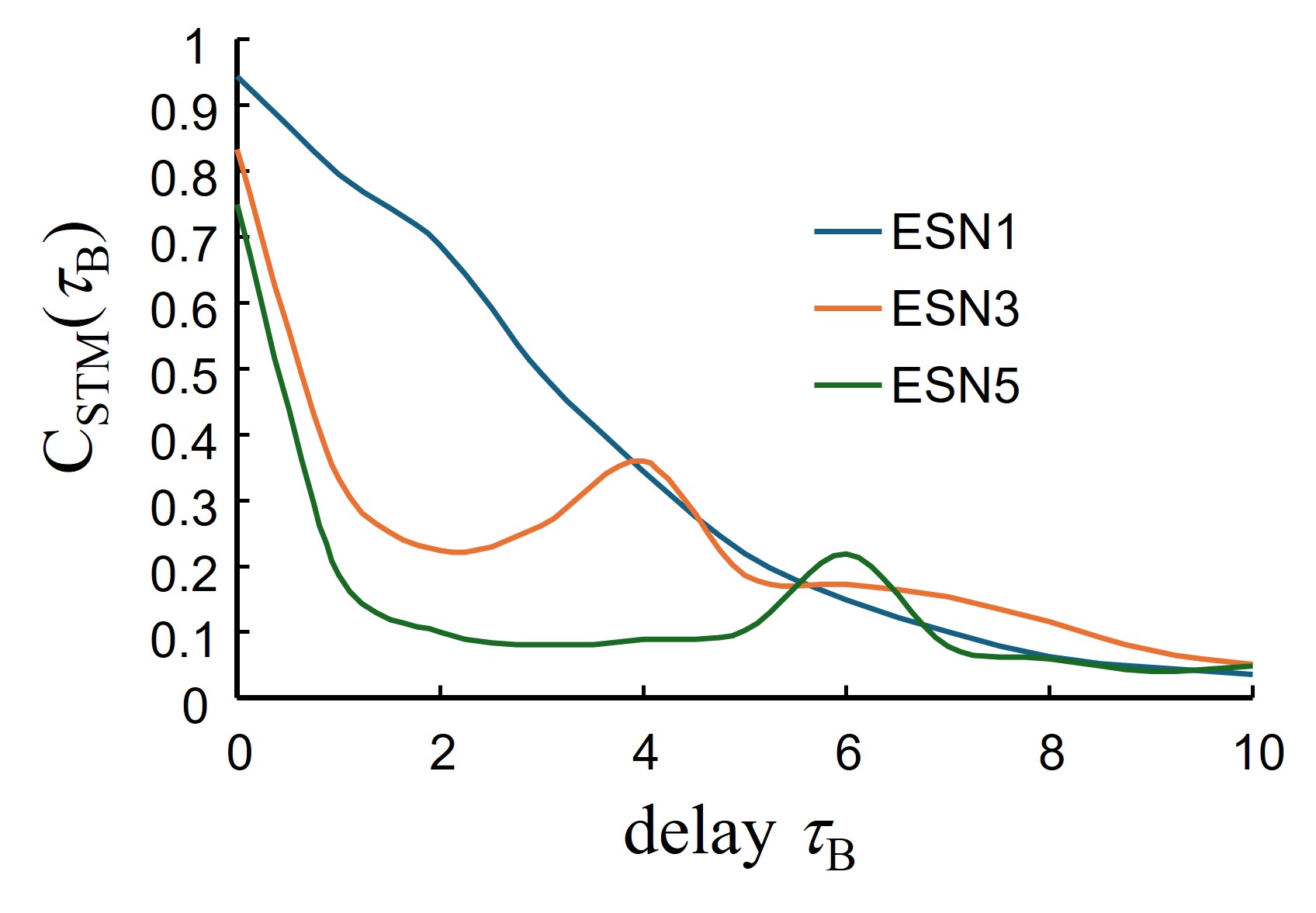}
\caption{The short-term memory test for the 
ESN for long correlations.
}
 \label{esnSTM}
\end{figure}

\section{Results}
Figure~\ref{fig2} shows the STM tests for types I and II, 
where $C_{\rm STM}(\tau_B)$ is called a STM capacity and given by the square of the correlation coefficient of $y_k$ and $\bar{y}_k$.
Because the number of the weights $w_i$ of the type I is larger 
than that of the type II, the type I shows better performance 
than the type II. 
Here, the linear cases for other patterns are not shown because they show smaller STM capacities. 
Figures~\ref{fig3a}-\ref{fig4b} are the examples 
of the NARMA tests for both the type I and II with linear and ring structures.
Tables I shows the results of NARMA tests for the types I and II.
NARMA2 and NARMA5 tasks show satisfactory results in both cases.
In the both cases, larger $\gamma$ cases show better performances.
This means that the dissipation process is important in QRC similar to the conventional RC.
Small $\gamma$ indicates that the system remembers the previous events,
but this long memory disturbs the newly inputted sequence.
When we look at the behavior of the soft silicone arm in Ref.~\cite{Nakajima1},
the soft silicone arm remembers only a couple of its previous steps.
The memory of the soft silicone arm disappears in the resistance of the water.
Comparing Figs.~\ref{fig3a} and \ref{fig3b} with Figs.~\ref{fig4a} and \ref{fig4b}
the typical waveform of the type II looks a little better than type I.

Our results suggest that the smaller dissipation 
induces inferior performance of our QRCs.
To confirm whether similar relation is satisfied even in ESN~\cite{ESN}, 
we modify the ESN function given by
\begin{equation}
x_{ki}=\tanh(\sum_j w_{ij}(x_{k-1,j}+
x_{k-3,j}+x_{k-5,j})+w_{{\rm in},i} s_k), 
\end{equation}
for the reservoir output $y_k=\sum_i w_{{\rm out},i} x_{ki}$
(ESN5).
Here, we take uniformly random $w_{ij}$ between 0 and 0.4. 
For ESN3 and ESN1, the terms after $w_{ij}$ 
are replaced by $x_{k-1,j}+x_{k-3,j}$, and $x_{k-1,j}$, 
respectively.
Figure \ref{esnSTM} and Table~\ref{f2} show the result of these long correlations of the ESN.
For longer correlation (ESN3, ESN5), the STM task shows poorer performance.
However, Tabel~\ref{f2} shows that there is no significant difference in NARMA tasks 
among ESN1, ESN3, ESN5. 
This is the same tendency as that of spin qubit system.
Thus, it could be said that the ESN given in Ref.~\cite{ESN} corresponds 
to a strong dissipation case of the physical system.

\begin{table}
\caption{\label{f2}NMSE in NARMA tasks for ESN of long correlations.}
\begin{tabular}{l|lll}
\hline
  &ESN1 &ESN3 &ESN5 \\ \hline
NARMA2 &9.85E-04& 9.64E-04& 2.22E-04\\
NARMA5 &1.98E-03& 3.59E-03& 1.61E-02\\
NARMA10 &2.74E-03& 4.00E-03& 1.47E-02\\
NARMA15 &3.48E-03& 3.49E-03& 7.45E-02\\
\hline
\end{tabular}
\end{table}

\section{Discussions}
The naive QRC method requires repeated initialization and measurement processes for ensemble averaging, 
which can potentially delay calculations. However, feedback mechanisms, 
such as those proposed by Kobayashi {\it et al.}~\cite{Kobayashi}, could be implemented to speed up QRC. Nonetheless, 
addressing this issue falls outside the scope of this study.

\begin{figure}
\centering
\includegraphics[width=8.2cm]{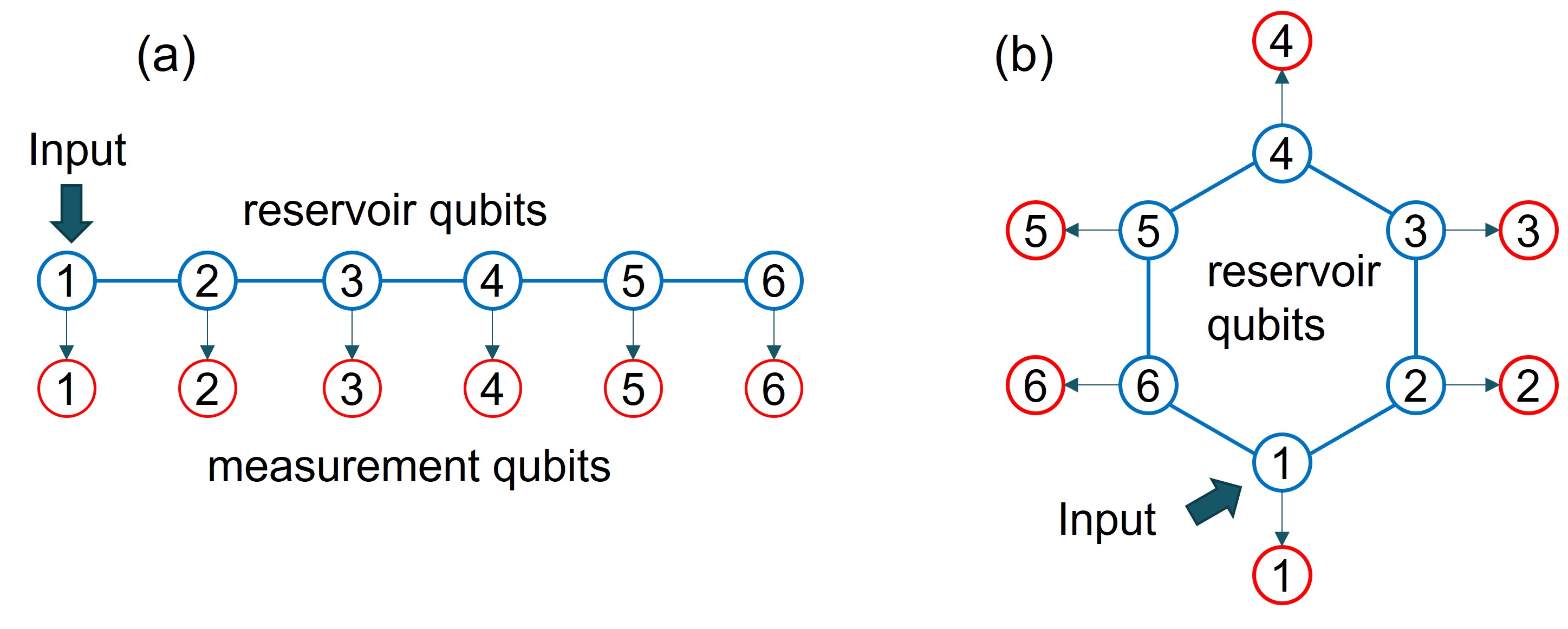}
\caption{The positions of the ancillary qubits (measurement qubits) for both the linear structure 
and the ring structure.
}
\label{PSB}
\end{figure}

In any case, measuring qubits will necessitate the use of ancillary qubits. 
For spin qubits, the Pauli spin blockade is a viable option~\cite{Ono}. 
Research by Yoneda {\it et al.} has demonstrated the high fidelity of the Pauli spin blockade~\cite{Yoneda}. 
In a QD system utilizing the Pauli spin blockade, each qubit is paired with an additional quantum dot, 
and the state of this additional QD is measured using single electron transistors.
The ancillary qubits are placed outside of the reservoir qubits (Fig.\ref{PSB}).
In the traditional gate-defined qubits~\cite{Intel,Dzurak,Vandersypen,Tarucha1}, 
the qubits and the wires are placed in the same plane.
In the case of the linear structure, both 
the reservoir qubits and ancillary qubits can be accessed from other side of
the qubit-qubit coupling. 
On the other hand,  in the ring structure, 
the access lines to the reservoir qubits are limited between the 
ancillary qubits.
Thus, from viewpoint of fabrication, the linear qubits are more feasible.

In this study, we carried out the STM task and NARMA task.
In order to estimate the qubit reservoir, other benchmark 
tasks such as the parity-check task, timer task and so on should be applied.
These are one of the future issues.
Here, the number of qubits are six. 
We have also estimated the STM task and NARMA task for $N_q=4$ case, but the results showed worse 
performance than those presented here (not shown here). 
Although the problem is its large computational costs,
the case of larger number of qubits will be a near-future problem.
 
\section{Conclusions}
We theoretically investigated a reservoir computing system 
using a spin qubit array.
Compared with the input encoding method where input sequence is exactly 
implemented to the first qubit, 
here, the decoherence prevails over all qubits.
Nevertheless, the NARMA tests show relatively good results, 
and it is shown that the QRC in spin qubit array is a promising 
as an application of spin qubits.
The detailed comparison with other RC system 
is a future issue.

\begin{acknowledgments}
We are grateful to K. Ono for the fruitful discussions we had.
This study was supported by JSPS KAKENHI Grant Number JP22K03497.
This study was also partially supported by the MEXT Quantum Leap Flagship Program (MEXT Q-LEAP; 
Grant Number JPMXS0118069228), Japan.
\end{acknowledgments}


\end{document}